\documentclass[12pt]{article}
\usepackage{graphics,psfig,epsfig}
   
\begin{document}


\def\beq             {\begin{equation}}
\def\eeq             {\end{equation}}
\def\beqd            {\begin{displaymath}}
\def\eeqd            {\end{displaymath}}
\def\baa             {\begin{array}}
\def\eaa             {\end{array}}
\def\beqaa           {\begin{eqnarray}}
\def\eeqaa           {\end{eqnarray}}
\def\beqaad          {\begin{eqnarray*}}
\def\eeqaad          {\end{eqnarray*}}
\def\btabu           {\begin{tabular}}
\def\etabu           {\end{tabular}}
\def\bfig            {\begin{figure}}
\def\efig            {\end{figure}}
\def\bce             {\begin{center}}
\def\ece             {\end{center}}

\def\ind             {\indent}
\def\noi             {\noindent}
\def\nn              {\nonumber}

\newcommand{\eq}[1]  {\mbox{eq.~(\ref{eq:#1})}}
\newcommand{\fig}[1] {\mbox{Fig.~\ref{fig:#1}}}   

\def\lbr             {\lbrack}
\def\rbr             {\rbrack}
\def\ti              {\tilde}
\def\q               {\bar}

\def\a               {\alpha}
\def\b               {\beta}
\def\d               {\delta}
\def\g               {\gamma}
\def\G               {\Gamma}
\def\l               {\lambda}
\def\t               {\theta}
\def\s               {\sigma}
\def\x               {\chi}

\def\de              {\partial}
\def\demu            {\partial_{\mu}}
\def\delr            {\!\stackrel{\leftrightarrow}{\partial^\mu}\!}

\def\sq              {\ti q}
\def\sqi             {\ti q_{i}^{}}
\def\sqL             {\ti q_{L}^{}}
\def\sqR             {\ti q_{R}^{}}
\def\sqe             {\ti q_{1}^{}}
\def\sqz             {\ti q_{2}^{}}
\def\sqp             {\ti q^{\prime}}
\def\qp              {q^{\prime}}

\def\st              {\ti t}
\def\sb              {\ti b}

\def\ch              {\ti \x^\pm}
\def\chp             {\ti \x^+}
\def\chm             {\ti \x^-}
\def\nt              {\ti \x^0}

\def\mV              {m_{\!V}}

\newcommand{\msq}[1]   {m_{\sq_{#1}}}
\newcommand{\mst}[1]   {m_{\st_{#1}}}
\newcommand{\msb}[1]   {m_{\sb_{#1}}}
\newcommand{\mch}[1]   {m_{\ti \x^\pm_{#1}}}
\newcommand{\mnt}[1]   {m_{\ti \x^0_{#1}}}

\def\dth             {\d\theta}

\def\sg              {{\ti g}}
\def\hg              {^{(g)}}
\def\hsg             {^{(\sg )}}
\def\hsq             {^{(\sq )}}
\def\msg             {m_{\sg}}

\def\tW              {\t_W}
\def\tsq             {\t_{\sq}}
\def\sth             {\sin\t}
\def\cth             {\cos\t}
\def\sthq            {\sin^2\t}
\def\cthq            {\cos^2\t}
\def\cst             {\cos\t_{\st}}
\def\csb             {\cos\t_{\sb}}

\def\Emiss           {E\llap/}

\def\lag             {{\cal L}}
\def\lum             {{\cal L}}
\def\R               {{\cal R}}
\def\Oas             {{\cal O}(\alpha_{s})}

\newcommand{\smaf}[2] {{\textstyle \frac{#1}{#2} }}
\def\onesq            {{\textstyle \frac{1}{\sqrt{2}} }} 
\def\onehf            {{\textstyle \frac{1}{2} }} 
\def\oneth            {{\textstyle \frac{1}{3} }}  
\def\twoth            {{\textstyle \frac{2}{3} }}
\def\onefo            {{\textstyle \frac{1}{4} }}
\def\forth            {{\textstyle \frac{4}{3} }}

\def\rzw             {\sqrt{2}}

\def\ra              {\rightarrow}

\def\BR              {\rm BR}
\def\gev             {\:{\rm GeV}}
\def\tev             {\:{\rm TeV}}
\def\pb              {${\rm pb}^{-1}$}

\def\leerz           {\hspace{1cm}\\}

\newcommand{\gsim}{\;\raisebox{-0.9ex}
           {$\textstyle\stackrel{\textstyle >}{\sim}$}\;}
\newcommand{\lsim}{\;\raisebox{-0.9ex}{$\textstyle\stackrel{\textstyle<}
           {\sim}$}\;}

\begin{titlepage}

\vspace*{-24mm}
\begin{flushright}
  UWThPh-1997-30 \\
  HEPHY-PUB 671/97 \\
  TGU-21 \\
  TU-526 \\
  hep-ph/9710286 \\[2mm]
\end{flushright}

\bce

\vspace{16mm}

{\Large {\bf SUSY-QCD corrections \\[2mm] 
             to stop and sbottom decays \\[2mm]
             into \boldmath{$W^{\pm}$} and \boldmath{$Z^0$} bosons \\[2mm]}}   

\vspace{12mm}

{\sc A. Bartl$^{\small 1}$, 
     H. Eberl$^{\small 2}$, 
     K. Hidaka$^{\small 3}$, } \\[2mm] 
{\sc S. Kraml$^{\small 2}$, 
     W. Majerotto$^{\small 2}$, 
     W. Porod$^{\small 1}$, 
     Y. Yamada$^{\small 4}$}

\vspace{10mm}

{\em (1) Institut f\"ur Theoretische Physik, Universit\"at Wien, 
         A-1090 Vienna, Austria} \\[1mm]
{\em (2) Institut f\"ur Hochenergiephysik, 
         \"Osterreichische Akademie der Wissenschaften, \\
         A-1050 Vienna, Austria}\\[1mm]
{\em (3) Department of Physics, Tokyo Gakugei University, Koganei, 
         Tokyo 184, Japan}\\[1mm]
{\em (4) Department of Physics, Tohoku University, Sendai 980-77, Japan}\\[1mm]

\vspace{1.3cm}

\begin{abstract}
We calculate the supersymmetric ${\cal O}(\alpha_s)$ QCD corrections 
to stop and sbottom decays into vector bosons within the Minimal 
Supersymmetric Standard Model. 
We give analytic formulae and perform a numerical analysis of these 
decays.
We find that SUSY--QCD corrections to the decay widths are typically 
$-5\%$ to $-10\%$ depending on the squark masses, squark mixing angles, 
and the gluino mass. 
\end{abstract}

\ece 
\end{titlepage}

\baselineskip 18pt

\section {Introduction}

In the Minimal Supersymmetric Standard Model (MSSM) \cite{haber} 
every quark has two scalar partners, the squarks $\sqL$ and $\sqR$.
In general $\sqL$ and $\sqR$ mix to form mass eigenstates $\sqe$ and $\sqz$ 
(with $m_{\sq_1} < m_{\sq_2}$), the size of the mixing being proportional 
to the mass of the corresponding quark $q$ \cite{ellis}. 
Therefore, the scalar partners of the top quark (stops) are expected to 
be strongly mixed so that one mass eigenstate $\st_1$ can be rather light 
and the other one $\st_2$ heavy. 
The sbottoms $\sb_L$ and $\sb_R$ may also considerably 
mix for large $\tan\b = v_2/v_1$ (where $v_1$ and $v_2$ are the 
vacuum expectation values of the two Higgs doublets).

Squark pair production in $e^+e^-$ annihilation including mixing 
has been studied at tree--level in Ref.~\cite{hikasa}, then 
including conventional QCD corrections in \cite{drees,beenakker}, 
and supersymmetric (SUSY) QCD corrections in Ref.~\cite{djouadi,helmut}. 
The cross section of squark production at hadron colliders (LHC and 
Tevatron) in next-to-leading order of SUSY--QCD was 
given in Ref.~\cite{hadro}.

While, quite naturally, most studies concentrated on $\st_1$ ($\sb_1$) 
production and decays, those of the heavier $\st_2$ ($\sb_2$) 
have been discussed much less \cite{porod,nlc,ecfa}.
These particles could be produced at the LHC or an $e^+e^-$ 
Linear Collider.
The decay patterns of the heavier mass eigenstates $\st_2$ ($\sb_2$) 
can be very complex due to the many possible open decay channels. 
There are the decays ($i,\,j = 1,2$; $k=1\ldots4$):
\beq
    \st_i \to t\,\nt_k,\; b\,\chp_j, \hspace{8mm}
    \sb_i \to b\,\nt_k,\; t\,\chm_j,
\eeq
and in case the mass splitting is large enough:
\beq
  \begin{array}{lcl}
    \st_i \to \sb_j\:W^+, &\hspace{3mm}& \sb_i \to \st_j\:W^-, \\
    \st_2 \to \st_1\,Z^0, &\hspace{3mm}& \sb_2 \to \sb_1\,Z^0,
  \end{array}
\label{eq:wzmodes}
\eeq
and
\beq
  \begin{array}{lcl}
    \st_i \to \sb_j\:H^+, &\hspace{4mm}& \sb_i \to \st_j\:H^-, \\
    \st_2 \to \st_1\:(h^0,\,H^0,\,A^0), &\hspace{4mm}& 
    \sb_2 \to \sb_1\:(h^0,\,H^0,\,A^0).
  \end{array}
\eeq

Stops and sbottoms can also decay strongly through $\st_i \to t\,\sg$ 
and $\sb_i \to b\,\sg$. If these decays are kinematically possible, 
they are important.
The SUSY--QCD corrections to squark decays into charginos 
and neutralinos have been calculated in \cite{chnt}.
The SUSY--QCD corrections to squark decays into 
Higgs bosons have been treated in \cite{higgsdec}, 
and those into gluino in \cite{gluino}. 
The decays to photon and gluon, 
$\sq_2\to\sq_1\g,\;\sq_1g$, which are absent at tree--level, 
are not induced by ${\cal O}(\a_s)$ SUSY--QCD corrections either.  
This is due to $\rm{SU}(3)_c\times\rm{U}(1)_{em}$ gauge invariance. 

In this paper we calculate the ${\cal O}(\alpha_s)$ 
SUSY--QCD corrections to the squark decays into $W^\pm$ and $Z^0$ bosons   
of \eq{wzmodes}. The stop decays into vector bosons can be dominant 
in the parameter region where 
(i) $m_t\,(A_t - \mu\,\cot\b)$, appearing in the mass matrix, 
    is large enough to give the necessary mass splitting, 
(ii) $M$ and $|\mu|$ are relatively large to suppress the decay modes 
to chargino, neutralino, and gluino, and 
(iii) the decays into Higgs particles are not so important (for 
instance, $m_A$ is large).
For sbottoms to decay into vector bosons instead of (i) it is 
necessary that $m_b\,(A_b - \mu\,\tan\b)$ is large enough to lead to a 
large mass splitting of $\sb_1$ and $\sb_2$.

In calculating the SUSY--QCD corrections we work in the on-shell 
renormalization scheme and use dimensional reduction \cite{siegel}, 
which preserves supersymmetry at least at two--loop level.
The gluon loop corrections are calculated in the Feynman gauge. 
The couplings of the squarks to vector bosons depend on the squark 
mixing angles for which an appropriate renormalization procedure 
is necessary. 
Here we use the renormalization prescription as 
introduced in \cite{helmut}, where it was applied to the case 
of $e^{+}e^{-} \to \sqi\bar{\sq}_{j}^{}$.

\section {Tree--level formulae and notation}

The current eigenstates $\sqL$ and $\sqR$ are related to their 
mass eigenstates $\sqe$ and $\sqz$ by 
\beq
  {\sqe \choose \sqz} = {\cal R}^{\sq}\,{\sqL \choose \sqR},
  \hspace{4mm}
  {\cal R}^{\sq} = \left(\baa{rr} \cth_{\sq} & \sth_{\sq} \\ 
                                 -\sth_{\sq} & \cth_{\sq} \eaa\right) 
  \hspace{5mm}
  (0 \leq \tsq < \pi ) .
\eeq

In the $(\sqe,\;\sqz)$ basis the squark interactions 
with $Z^0$ and $W^\pm$ bosons are given by:
\beqaa
  {\cal L} &=& 
  -\frac{ig}{\cos\tW}\, Z_\mu\, 
      (\R^{\sq}_{i1}\R^{\sq}_{j1} C_L + \R^{\sq}_{i2}\R^{\sq}_{j2} C_R)\, 
      \sq_j^\dagger \delr \sqi \nn\\[2mm]
  & &    
  -\frac{ig}{\rzw}\, \R^{\ti t}_{i1} \R^{\ti b}_{j1}\, 
       W^-_\mu\, \ti b_j^\dagger \delr \ti t_i^{} 
  -\frac{ig}{\rzw}\, \R^{\ti b}_{i1} \R^{\ti t}_{j1}\, 
       W^+_\mu\, \ti t_j^\dagger \delr \ti b_i^{},
\label{eq:Lint}
\eeqaa
where $C_{L,R} = I_{3L,R}^{q} - e_q \sin^2\tW$ with $I_3^q$  
the third component of the weak isospin and $e_q$ the electric charge. 
Here it is assumed that the $(3,3)$ element of the super--CKM matrix 
$\ti K_{33} = 1$.

At tree--level the amplitude of a squark decay into 
a $W^\pm$ or $Z^0$ boson has the general form
\beq
  {\cal M}_0(\sq^\a_i \to \sq^\b_j V) = 
  -ig\,c_{ijV}\, (k_1 + k_2)^\mu\, \epsilon_\mu^* (k_3),
\label{eq:Mtree}  
\eeq
with $k_1$, $k_2$, and $k_3$ the four--momenta of $\sq^\a_i$, 
$\sq^\b_j$, and the vector boson $V$ ($V = W^\pm, Z^0$), respectively 
(Fig.~1a). $\a$ and $\b$ are flavor indices. 
In the following we define $m_i = m_{\sq^\a_i}$, $m_j = m_{\sq^\b_j}$, 
$\R_{ik} = \R_{ik}^{\sq^\a}$, $\R_{jk} = \R_{jk}^{\sq^\b}$ for simplicity. 
Moreover, we shall use primes to explicitly distinguish between different 
flavors.
 
With this notation the $\sq^\a_i$-$\sq^\b_j$-$V$ couplings 
$c_{ijV}$ are:
\beqaa
c_{ijZ} 
  &=& \smaf{1}{\cos\tW}\,(\R_{i1}\R_{j1} C_L + \R_{i2}\R_{j2} C_R) 
  \nn\\[2mm]
  &=& \smaf{1}{\cos\tW}\, 
    \left( \begin{array}{cc}
      I_{3L}^{q}\,\cos^2\tsq - e_{q} \sin^2\tW 
        & -\onehf\, I_{3L}^{q}\,\sin 2\tsq \\
    -\onehf\, I_{3L}^{q}\,\sin 2\tsq 
        & I_{3L}^{q}\,\sin^2\tsq - e_{q} \sin^2\tW
    \end{array} \right)_{ij} , \\[4mm]
c_{ijW} 
  &=& \smaf{1}{\rzw}\, \R_{i1}\R'_{j1}  
  = \smaf{1}{\rzw}\, 
    \left( \begin{array}{rr}
       \cos\tsq \cos\t_{\sq'} & -\cos\tsq \sin\t_{\sq'} \\
      -\sin\tsq \cos\t_{\sq'} &  \sin\tsq \sin\t_{\sq'}
    \end{array} \right)_{ij} .
\eeqaa 

The tree--level decay width can thus be written as \cite{porod}
\beq
  \Gamma^{0} (\sq^\a_i \to \sq^\b_{j} V) =
  \frac{  g^2\, (c_{ijV})^2\, \kappa^3 (m_i^2, m_j^2, \mV^2)
       }{ 16\pi\, \mV^2\, m_i^3 } ,
\label{eq:Gtree}
\eeq

with $\kappa (x,y,z) = (x^{2}+y^{2}+z^{2}-2xy-2xz-2yz)^{1/2}$. 

Note that the decay $\sq_{L(R)}^{} \to \sq_{R(L)}^{}\, Z^0$ does not 
occur at tree--level, and $\sqi \to \sq'_j\, W^\pm$ is not allowed 
at tree--level if one of the squarks is a pure ``right'' state.

\section {SUSY--QCD corrections}

The ${\cal O}(\a_{s})$ loop corrected decay amplitude is obtained by 
the shift:  
\beq
c_{ijV}\to c_{ijV} + \d c_{ijV}^{(v)} + \d c_{ijV}^{(w)} 
+ \d c_{ijV}^{(c)}, 
\eeq
in \eq{Mtree}. 
The superscript $v$ denotes the vertex correction (Figs.~1\,b--f), 
$w$ the wave function correction (Figs.~1\,g--i), 
and $c$ the counterterm to the couplings. 
Thus the ${\cal O}(\a_{s})$ corrected decay width $\G$ is given by 
\beq
  \G = \G^{0} + \d\G^{(v)} + \d\G^{(w)} + \d\G^{(c)} + \d\G_{real} 
  \label{eq:gencorr}
\eeq
where 
\beq
  \d\G^{(a)} = 
  \frac{g^2\,\kappa^3(m_i^2,m_j^2,\mV^2)}{8\pi\,\mV^2\,m_i^3}\;
  c_{ijV}\:\mbox{Re}\left\{\d c_{ijV}^{(a)}\right\}  
  \hspace{6mm}(a = v,w,c). 
\label{eq:deltaG}
\eeq
$\d\G_{real}$ is the correction due to real gluon emission 
(Figs.~1j--$\ell$) which is included in order to cancel the 
infrared divergence. 
The total correction can be decomposed into a gluon (exchange and 
emission) contribution $\d\G^{(g)}$, and a gluino--exchange contribution 
$\d\G^{(\sg)}$: $\G = \G^{0} + \d\G^{(g)} + \d\G^{(\sg)}$. 
The contribution from squark loops (Figs. 1\,f,i) cancels out in 
our renormalization scheme. 

\subsection{Vertex corrections}
                 
The vertex corrections stem from the five diagrams shown in Figs.~1\,b--f. 
The gluon--exchange graphs of Figs.~1b--d yield 
\beqaa
  \d c_{ijV}^{(v,g)} 
  &=& -\frac{\a_s}{3\pi}\: c_{ijV} \left[ \,
    B_0(m_i^2,\l^2,m_i^2) + B_0(m_j^2,\l^2,m_j^2) \right. \nn\\
  & & \hspace{40mm} \left.
    -\,2\,(m_i^2 + m_j^2 - \mV^2)\,(C_0 + C_1 + C_2)\, \right] .
\eeqaa
$B_0$ and $C_{0,1,2}$ are the standard two-- and three--point 
functions \cite{pave} for which we follow the conventions of \cite{denner}. 
Here we use the abbreviation 
$C_{m} = C_{m}(m_i^2, \mV^2, m_j^2; \l^{2}, m_i^2, m_j^2)$. 
A gluon mass $\l$ is introduced to regularize the infrared divergence. 

The gluino--exchange contribution, Fig.~1e, gives 
\vspace{1mm} 
\beqaa
  \d c_{21Z}^{(v,\sg)} 
  &=& -\,\frac{\a_s}{3\pi\cos\t_W}\: \Big\{
    I_{3L}^q\,\big[\,2\,\msg^2\,C_0 + m_{\sq_2}^2\,C_1 + m_{\sq_1}^2\,C_2 
    +(\msg^2 - m_q^2)\,(C_1+C_2) + B_0\,\big]\, \sin 2\tsq \nn\\         
  & & \hspace{26mm} 
    +\,2\,\msg\,m_q\,(I_{3L}^q - 2e_q\sin^2\t_W)\,
                     (C_0+C_1+C_2)\,\cos 2\tsq \Big\} ,
\eeqaa
with 
$C_{m} = C_{m}(m_{\sq_2}^2, m_{\!Z}^2, m_{\sq_1}^2; \msg^{2}, m_q^2, m_q^2)$ 
and $B_0 = B_0(m_{\!Z}^2, m_q^2, m_q^2)$, 
for the decay $\sqz \to \sqe\, Z^0$, and 
\beqaa
  \d c_{ijW}^{(v,\sg)} 
  &=& -\,\frac{\sqrt{2}}{3}\, \frac{\a_s}{\pi}\: \Big\{
    \msg\,(C_0 + C_1 + C_2)\,
          (m_q\,\R_{i2}^{}\,\R'_{j1} + m_{q'}\,\R_{i1}^{}\,\R'_{j2}) \nn\\
  & & \hspace{25mm}
  -\,\big[\,m_i^2\,C_1 + m_j^2\,C_2 + \msg^2\,(2C_0+C_1+C_2) + B_0\,\big]\,
     \R_{i1}^{}\,\R'_{j1} \nn\\
  & & \hspace{25mm} -\, m_q\,m_{q'}\,(C_1 + C_2)\,\R_{i2}^{}\,\R'_{j2}\, 
  \Big\}
\eeqaa
with $C_{m} = C_{m}(m_i^2, m_{\!W}^2, m_j^2; \msg^{2}, m_q^2, m_{q'}^2)$
and $B_0 = B_0(m_{\!W}^2, m_q^2, m_{q'}^2)$, 
for the decay $\sqi\to\sq'_j\, W^\pm$. 
The squark loop of Fig.~1f does not contribute because it is 
proportional to the four--momentum of the vector boson.
The total vertex correction is thus given by: 
\beq
  \d c_{ijV}^{(v)} = 
  \d c_{ijV}^{(v,g)} + \d c_{ijV}^{(v,\sg)}.
\eeq

\subsection{Wave--function correction}

The wave--function correction is given by
\beq 
  \d c_{ijV}^{(w)} = 
  \onehf \left[ 
    \d\ti Z_{ii}(\sq_i^\a) + \d\ti Z_{jj}(\sq_j^\b) \right] c_{ijV}
  + \d\ti Z_{ik}(\sq_i^\a)\,c_{kjV} + \d\ti Z_{jl}(\sq_j^\b)\,c_{ilV},  
  \quad (i\neq k, j\neq l).   
\eeq

$\ti Z_{nm}(\sq_n)$ are the squark wave--function renormalization constants. 
They stem from gluon, gluino, and squark loops (Figs.~1\,g--i) and 
are given by:
\beq
  \delta \ti Z_{nn}^{(g,\sg)}(\sq_n) = 
    - \mbox{Re}\left\{\dot\Sigma_{nn}^{(g,\sg)}(\msq{n}^{2}) \right\}\,, 
  \quad
  \delta \ti Z_{nn'}^{(\sg,\sq)}(\sq_n) =
  -\,\frac{\mbox{Re}\left\{\Sigma_{nn'}^{(\sg,\sq)}(\msq{n}^{2})\right\} 
         }{\msq{n}^2 - \msq{n'}^2} \,, 
  \quad n \neq n' ,
\eeq
with $\dot\Sigma_{nn} (m^2) = 
\partial\Sigma_{nn} (p^2)/\partial p^2 |_{p^2=m^2}$. 
(The gluon loop due to the $\sq\sq gg$ interaction gives no 
contribution because it is proportional to the gluon mass $\l\to 0$.)
The squark self--energy contribution
due to gluon exchange (Fig.~1g) is
\beq
  \dot\Sigma_{nn}^{(g)} (\msq{n}^2) = - \frac{2}{3}\frac{\alpha_s}{\pi} 
  \left[
     B_{0}(\msq{n}^{2}, 0, \msq{n}^{2}) 
     + 2\msq{n}^{2} \dot B_{0} (\msq{n}^{2}, \l^{2}, \msq{n}^{2})
  \right] ,
\eeq
and that due to gluino exchange (Fig.~1h)  is
\beqaa  
  \dot\Sigma_{nn}^{(\sg)} (\msq{n}^2) &=& \frac{2}{3}\frac{\a_{s}}{\pi} 
  \left[ B_0(\msq{n}^2, \msg^2,m_q^2) + (\msq{n}^2-m_q^2-\msg^2)\, 
         \dot B_0 (\msq{n}^2, \msg^2, m_q^2) \right.\nn\\
  & & \left.\hspace*{12mm} 
    -\, 2 m_q\msg (-1)^n\sin 2\tsq\, \dot B_0 (\msq{n}^2, \msg^2,m_q^2) 
  \right] ,
\eeqaa
\beq
  \Sigma_{12}^{(\sg)}(\msq{n}^2) = \Sigma_{21}^{(\tilde g)}(\msq{n}^2) =
  \frac{4}{3}\frac{\alpha_{s}}{\pi}\, \msg m_q \cos 2 \tsq \,
  B_0(\msq{n}^2, \msg^2,m_q^2).
\eeq
The four--squark interaction (Fig.~1i)  gives
\beq
\Sigma_{12}\hsq (\msq{n}^2) = \Sigma_{21}\hsq (\msq{n}^2) = 
  \frac{\alpha_s}{6\pi}\sin 4\tsq
  \left[ A_{0}(m_{{\sq}_2}^2) - A_{0}(m_{{\sq}_1}^2) \right] ,
\eeq
where $A_{0}(m^2)$ is the standard one--point function in the 
convention of \cite{denner}. 
Note that $\Sigma_{nn'}\hsq (\msq{1}^2) = \Sigma_{nn'}\hsq (\msq{2}^2)$.

\subsection{On-shell renormalization of squark mixing angles}

It is necessary to renormalize the squark mixing angles $\tsq$ 
by adding appropriate counterterms to obtain ultraviolet finite 
decay widths:
\beq
  \d c_{21Z}^{(c)} = - \smaf{1}{\cos\t_W}I_{3L}^{q} \cos 2\tsq\, \d\tsq ,
\label{tcounter}
\eeq
\beqaa
  \d c_{ijW}^{(c)}  
  &=& \smaf{1}{\rzw}\, \left[
    \left( \begin{array}{rr}
      -\sin\tsq \cos\t_{\sq'} & \sin\tsq \sin\t_{\sq'} \\
      -\cos\tsq \cos\t_{\sq'} & \cos\tsq \sin\t_{\sq'}
    \end{array} \right) \d\tsq \right.\nn \\[2mm]
  & & \hspace{30mm} + \left. \left( \begin{array}{rr}
      -\cos\tsq \sin\t_{\sq'} & -\cos\tsq \cos\t_{\sq'} \\
       \sin\tsq \sin\t_{\sq'} &  \sin\tsq \cos\t_{\sq'}
    \end{array} \right) \d\t_{\sq'} \right]_{ij} .
\eeqaa
For the definition of the on-shell mixing angles $\tsq$ 
we follow the procedure given in Ref.~\cite{helmut}. 
The counterterm $\d\tsq$ is fixed such that it cancels the 
off-diagonal part of the squark wave function corrections 
to the cross section of $e^+e^-\to\sq_1^{}\bar{\sq}_2$.
$\d\tsq=\delta\tsq\hsq+\delta\tsq\hsg$ is then given by 
\beq  
  \delta\tsq\hsq = \frac{\alpha_s}{6\pi} \,
  \frac{\sin 4\tsq}{\msq{1}^2 - \msq{2}^2}
  \left[ A_{0}(\msq{2}^2) - A_{0}(\msq{1}^2) \right] ,
\eeq
\beq  
  \delta\tsq\hsg = \frac{\alpha_s}{3\pi}
  \frac{m_{\sg} m_q}{I_{3L}^q (m_{\sq_1}^2 - m_{\sq_2}^2)}
  \left[ B_{0}(m_{\sq_2}^2,m_{\sg}^2,m_q^2)\,\ti v_{11} -
         B_{0}(m_{\sq_1}^2,m_{\sg}^2,m_q^2)\,\ti v_{22}\right] ,
\eeq

\noi
with 
$\ti v_{11}= 4(I_{3L}^{q} \cos^2 \tsq - e_{q}\sin^2\t_W^{})$ and 
$\ti v_{22}= 4(I_{3L}^{q} \sin^2 \tsq - e_{q}\sin^2\t_W^{})$. 
Obviously, for the decay $\sq_2^{}\to\sq_1^{}\,Z^0$, 
the counterterm (\ref{tcounter}) 
completely cancels the off-diagonal wave function corrections 
(Figs.~1h ($i\not=n$) and 1i ($i\not=k$)). 
In case of $\sqi \to \sq_j^{\prime}\,W^\pm$ the contribution of the squark 
loop, Fig. 1i, is cancelled. Thus, in both cases the total squark loop 
contribution to the correction is zero, $\d\G^{(\sq)}\equiv 0$.

\subsection{Real gluon emission}

In order to cancel the infrared divergence we include 
real (hard and soft) gluon emission: 
$\d\G_{real} = \G(\sq_i^\a \to \sq_j^\b\,V\,g)$   
(Figs.~1j, 1k, and 1$\ell$). The width is
\beq
  \d\G_{real} 
  = \frac{g^2\,c_{ijV}^2\,\a_s}{3\pi^2 m_i}\:\left\{ 
  2 I - \frac{\kappa^2}{m_V^2}
  \left[\, I_0 + I_1 + m_i^2\,I_{00} + m_j^2\,I_{11} 
               + (m_i^2 + m_j^2 - m_V^2)\,I_{01} \right] 
  \right\} .
\eeq
Again, $\kappa = \kappa(m_i^2,\,m_j^2,\,m_V^2)$. 
The phase space integrals $I$, $I_{n}$, and $I_{nm}$ 
have $(m_i, m_j, m_V)$ as arguments. 
Their explicit forms are given in \cite{denner}.

We have checked explicitely that the corrected decay width, 
eq.~(\ref{eq:gencorr}), is ultraviolet and infrared finite. 

\section{Numerical results and discussion}

In general, the stop and sbottom sectors are determined by the 
soft SUSY breaking parameters ($M_{\ti Q}$, $M_{\ti U}$, 
and $M_{\ti D}$), the trilinear couplings ($A_{t}$ and $A_{b}$),  
$\mu$, and $\tan\b$, which all enter the squark mass matrices. 
In order to show the importance of the $\sq_2^{}\to\sq_1^{} Z^0$ 
and $\sqi\to\sq^\prime_j W^\pm$ decays, we plot in Fig. 2 the branching 
ratios of these modes as a function of $\mu$ for 
$M_{\ti Q}=500\gev$, $M_{\ti U}=444\gev$, $M_{\ti D}=556\gev$, and 
$A_t=A_b=500\gev$. For the SU(2) gaugino mass we take $M=200\gev$, and 
for the mass of the pseudoscalar Higgs $m_A=200$ GeV   
(for the U(1) gaugino mass $M^\prime$ and the gluino mass we use the GUT 
relations $M^\prime =\frac{5}{3}\,M\tan^2\t_W$ and $\msg=\a_s/\a_2\,M$). 
Fig.~2a (2b) is for $\tan\b=2$ (40). We see that the squark decays into 
vector bosons can have very large branching ratios under the conditions 
(i), (ii), and (iii) given in the introduction.   

We now turn to the numerical analysis of the ${\cal O}(\a_{s})$ 
SUSY--QCD corrected decay widths. 
As the squark couplings to vector bosons depend only on 
the squark mixing angles, we use the on-shell squark masses $\msq{1,2}$ 
and mixing angles $\tsq$ ($0\le\tsq<\pi$) as input parameters. 
Moreover, we take $m_{t} = 175$ GeV, $m_{b} = 5$ GeV, 
$m_{Z} = 91.2$ GeV, $\sin^{2}\t_{W} = 0.23$, 
$\a(m_{Z}) = 1/128.87$, and
$\a_{s}(m_{Z}) = 0.12$. 
For the running of $\a_{s}$ we use $\a_{s}(Q^{2}) = 
12\pi/[(33 - 2\,n_{f})\,\ln (Q^{2}/\Lambda^{2}_{n_{\!f}})]$ 
with $n_{\!f}$ the number of flavors. 


We first discuss the decay $\st_2\to\st_1 Z^0$. 
Figure~3 shows the tree--level and the ${\cal O}(\a_{s})$ 
SUSY--QCD corrected widths of this decay as a function 
of the lighter stop mass $\mst{1}$, for $\mst{2}=650\gev$, $\cst=-0.6$, 
and $\msg=500\gev$. 
SUSY--QCD corrections reduce the tree--level width by $-11.7\%$ to $-6.8\%$ 
in the range of $\mst{1} = 80$ to 508~GeV.  
It is interesting to note that the gluonic correction decreases 
quickly with increasing $\mst{1}$ while the correction due to gluino 
exchange varies only little with $\mst{1}$. 
In our example, $\d\G^{(g)}/\G^0=-4.5\%$ and $\d\G^{(\sg)}/\G^0=-7.2\%$ 
at $\mst{1} = 80\gev$, whereas at $\mst{1} = 550\gev$ 
$\d\G^{(g)}/\G^0 \simeq 0\%$ and $\d\G^{(\sg)}/\G^0=-6.8\%$.\\ 
The dependence on the stop mixing angle is shown in Fig.~4.
Fig.~4a shows the tree--level width together with 
the ${\cal O}(\a_{s})$ corrected width of $\st_2\to\st_1 Z^0$ 
as a function of $\cst$ for $\mst{1}=200\gev$, $\mst{2}=650\gev$, 
and $\msg=500\gev$. 
Assuming $M_{\ti U} < M_{\ti Q}$, as suggested by SUSY--GUT, 
the stop mixing angle is varied in the range 
$-\frac{1}{\sqrt{2}} < \cst < \frac{1}{\sqrt{2}}$. 
With the $\st_1$-$\st_2$-$Z^0$ coupling proportional to 
$\sin 2\t_{\st}$ the decay width has maxima at 
$\cst=\pm \frac{1}{\sqrt{2}}$ (maximal mixing) and vanishes 
at $\cst=0$. 
In Fig.~4b we plot the relative correction $\d\G/\G^0$ 
for $m_{\st_{\{1,2\}}}\! = \{200,650\}\gev$ and $\msg=500$ and 
1000 GeV.
As the gluonic correction has the same $\t_{\st}$ dependence as 
the tree--level width, $\d\G^{(g)}/\G^0 = -3\%$ in our example, 
the $\t_{\st}$ dependence in Fig.~4b comes only from the correction due to 
gluino exchange. 
For $\msg=500\gev$ and $\cst \lsim -0.1$ ($\cst \gsim 0.1$) the correction 
is $-10\%$ to $-12\%$ ($-3\%$ to $-5\%$); for $\msg=1\tev$ and 
$|\cst| \gsim 0.1$ the correction is $-2\%$ to $-5\%$. 
Approaching $\cst=0$, where $\st_{1(2)}^{}=\st_{R(L)}^{}$, 
$\d\G/\G^0$ diverges due to the vanishing tree--level coupling 
$c_{21Z}$ while $\d c_{21Z}\ne 0$. 
In this case the decay width becomes of ${\cal O}(\a_s^2)$. 
Note, however, that the appearance of this divergence, as well as the 
condition $\cst=0$, is renormalization scheme dependent. \\
Taking a closer look on the gluino mass dependence we find that  
the gluino decouples slowly; e.g. for $\mst{1}=200\gev$, 
$\mst{2}=650\gev$, and $\cst=-0.6$ the gluino contribution 
to the correction 
$\d\G^{(\sg)}/\G^0 = -2.6\%,\; -0.8\%,\; -0.4\%$ for 
$\msg = 600\gev,\; 1\tev,\; 1.5\tev$, respectively. 
On the other hand, the size of the gluino contribution quickly increases 
when approaching the $\st_2^{}\to t\sg$ threshold: For $\msg=480\gev$ 
$\d\G^{(\sg)}/\G^0 = -17.2\%$, and for $\msg=476\gev$ 
$\d\G^{(\sg)}/\G^0 = -38.8\%$.
For $\msg < \mst{2}-m_t$ the gluino correction becomes positive. In 
our example it reaches the maximum at $\msg=270\gev$ with 
$\d\G^{(\sg)}/\G^0 = 3.4\%$. In general, the dependence on the gluino 
mass near the threshold is less pronounced for $\cst>0$. \\
The decay $\sb_2\to\sb_1 Z^0$ can be important for large $\tan\b$ 
(see Fig.~2b). 
The SUSY--QCD corrections to this decay behave similarly to those to 
$\st_2\to\st_1 Z^0$. 
However, the corrections due to gluino exchange are smaller because of 
the smaller bottom quark mass and thus the dependece of $\d\G/\G^0$ 
on the sbottom mixing angle is very weak.


Let us now turn to the squark decays into $W^\pm$ bosons. 
Here we discuss two special cases: \\
(i) $\sb_1$ and $\sb_2$ decaying into a relatively light $\st_1$ plus 
$W^-$ for small sbottom mixing (small $\tan\b$ scenario). 
In this case the mass difference of $\sb_1$ and $\sb_2$ 
is expected to be rather small and thus the decays 
$\sb_2\to\sb_1\, (Z^0\!, h^0\!, H^0\!, A^0)$ 
should be kinematically suppressed or even forbidden. \\
(ii) A heavy $\st_2$ decaying into a relatively light $\sb_1$ plus 
$W^+$ for large sbottom mixing (large $\tan\b$ scenario). \\
Note, however, that in a combined treatment of both the stop and the 
sbottom sectors there is a constraint among the 
parameters $\mst{1}$, $\mst{2}$, $\t_{\st}$, $\msb{1}$, $\msb{2}$, 
and $\t_{\sb}$ (for a given value of $\tan\b$) 
due to $\rm{SU(2)_L^{}}$ gauge symmetry \cite{nlc,mQproblem}. 
Therefore one of these parameters is fixed by 
the others. Here one has to take into account that in the 
on-shell scheme $M_{\ti Q}^2(\st) \not= M_{\ti Q}^2(\sb)$  
at ${\cal O}(\a_{s})$, see Ref.~\cite{mQproblem}. \\
The decay widths of $\,\sb_{1,2}\to\st_1 W^-$ are shown in Fig.~5a as a 
function of the stop mixing angle, for $\msb{1}=500\gev$, 
$\msb{2}=520\gev$, $\csb=-0.9$, $\mst{1}=200\gev$, $\msg=520\gev$, 
and $\tan\b=2$. 
The value of $\mst{2}$ is determined by the other 
parameters as discussed above. 
Hence $\mst{2}$ varies from 533 GeV to 733 GeV depending on the stop 
mixing angle in Fig.~5a.
(Here we have also checked that the squark parameters do not cause 
too large $A_t$ or $A_b$ to avoid color breaking minima.)
Despite the larger phase space for the $\sb_{2}$ decay, the width of 
$\sb_{2}\to\st_1 W^-$ is smaller than that of $\sb_{1}\to\st_1 W^-$ 
because the $W$ couples only to the ``left'' components of the squarks 
(note that $\sb_{1}\sim\sb_{L}^{}$ and $\sb_{2}\sim\sb_{R}^{}$ 
for $\csb=-0.9$).
Fig.~5b shows the relative correction 
$\d\G/\G^0$ for the same squark parameters as in Fig.~5a. 
For $\msg=520\gev$ and $|\cst|\gsim 0.1$ SUSY--QCD corrections 
are $-11\%$ to $+4\%$. 
For $\msg=1\tev$ the effects are weaker, i.e. about $-5\%$ 
to $-1\%$. 
Again, there is a ``singularity'' at $\cst=0$ where the tree--level 
$\st_1^{}$-$\sb_i^{}$-$W^\pm$ coupling vanishes.  
The dependence of $\d\G/\G^0$ on the sbottom mixing angle 
is, in general, much weaker than that on the stop mixing angle 
(apart from a singularity for the $\sb_i^{}$ being a pure $\sb_R^{}$). 
The overall dependence on the gluino mass is in general similar to that 
of the $\st_2^{}\to\st_1^{}Z^0$ decay. However, the enhancement 
effect of the threshold at $\msg = \msb{i}-m_b$ is less pronounced. \\
An example for large $\tan\b$ is shown in Fig.~6.  
In Fig.~6a we plot the tree--level and the SUSY--QCD corrected widths 
of $\st_2\to\sb_1 W^+$ as a function of $\cst$ for $\mst{1}=300\gev$, 
$\mst{2}=650\gev$, $\msb{1}=380\gev$, $\csb=-0.8$, $\msg=500\gev$, 
and $\tan\b=40$. 
$\msb{2}$ is calculated from the other parameters and thus 
varies from 615 GeV to 918 GeV.  
As expected, the decay width is maximal for $\st_2=\st_L$ and vanishes 
for $\st_2=\st_R$. In the example chosen, SUSY--QCD corrections are 
$-2.4\%$ to $-4.7\%$ for $\msg=500\gev$ and 
about $-1\%$ to $-1.5\%$ for $\msg=1\tev$ as 
can be seen in Fig.~6b, where $\d\G/\G^0$ is shown 
as a function of $\cst$ for the same squark parameters as given above, 
and $\msg=500\gev$ and $1\tev$.
Again, there is almost no dependence on $\csb$ apart from the 
singularity at $\csb=0$, i.e. $\sb_1^{}=\sb_R^{}$.

\section{Summary}

We have calculated the ${\cal O}(\a_{s})$ supersymmetric QCD 
corrections to squark decays into vector bosons in the on-shell 
renormalization scheme using dimensional reduction.  
In particular, we have discussed examples for the decays 
$\st_2\to\st_1 Z^0$, $\st_2\to\sb_1 W^+$, and $\sb_{1,2}\to\st_1 W^-$. 
We have found that the correction $\d\G/\G^0$ is typically of 
the order $-5\%$ to $-10\%$, depending on the squark masses, squark mixing 
angles, and $\msg$. Near the $\sq\to q\sg$ threshold the correction 
can also exceed $-10\%$. 
It has also turned out that the gluino decouples slowly. 
Moreover, the gluino--exchange corrections alter the $\tsq$ dependence 
of the tree--level widths. 
For squark mixing angles where the decay width vanishes at 
tree--level, the gluino corrections do not vanish and may 
lead to non-zero widths of ${\cal O}(\a_{s}^2)$. 

\section*{Acknowledgements}

The work of A.B., H.E., S.K., W.M., and W.P. was supported by 
the ``Fonds zur F\"orderung der wissenschaftlichen Forschung'' 
of Austria, project no. P10843--PHY.

\clearpage


\setlength{\unitlength}{1mm}
\begin{center}


\clearpage

\begin{picture}(125,210)
\put(0,15){\mbox{\psfig{file=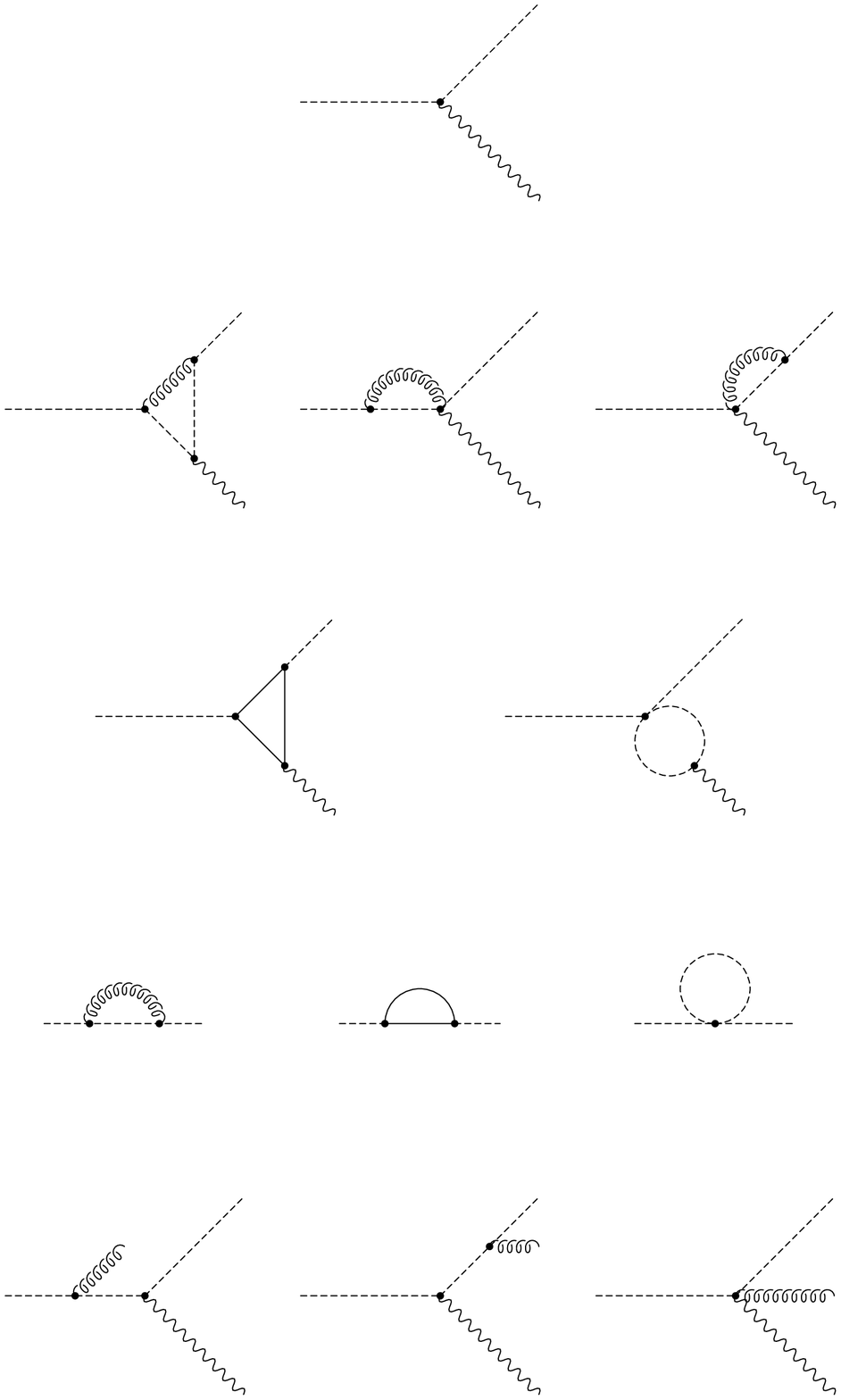,height=195mm}}}
\put(57,181){\makebox(0,0)[t]{\bf{(a)}}}
\put(39,196.5){\makebox(0,0)[r]{$\sq_i^\a$}}
\put(77,210){\makebox(0,0)[l]{$\sq_j^\b$}}
\put(77,182){\makebox(0,0)[l]{$V$}}
\put(51,193){\makebox(0,0)[t]{$k_1$}}
\put(62.5,203.5){\makebox(0,0)[b]{$k_2$}}
\put(63,188){\makebox(0,0)[t]{$k_3$}}
\put(49,194.5){\vector(1,0){4.3}}
\put(64,202){\vector(1,1){3.6}}
\put(63,190){\vector(1,-1){3.6}}


\put(-0.5,154){\makebox(0,0)[r]{$\sq_i^\a$}}
\put(35,168){\makebox(0,0)[l]{$\sq_j^\b$}}
\put(35,139){\makebox(0,0)[l]{$V$}}
\put(21.5,160.5){\makebox(0,0)[r]{$g$}}
\put(21.5,147.5){\makebox(0,0)[r]{$\sq_i^\a$}}
\put(28,154){\makebox(0,0)[l]{$\sq_j^\b$}}
\put(16,140){\makebox(0,0)[t]{\bf{(b)}}}

\put(40,154){\makebox(0,0)[r]{$\sq_i^\a$}}
\put(76,168){\makebox(0,0)[l]{$\sq_j^\b$}}
\put(76,139){\makebox(0,0)[l]{$V$}}
\put(58,149){\makebox(0,0)[r]{$\sq_i^\a$}}
\put(56.5,161){\makebox(0,0)[bc]{$g$}}
\put(57,140){\makebox(0,0)[t]{\bf{(c)}}}

\put(81,154){\makebox(0,0)[r]{$\sq_i^\a$}}
\put(117,168){\makebox(0,0)[l]{$\sq_j^\b$}}
\put(117,139){\makebox(0,0)[l]{$V$}}
\put(107,155){\makebox(0,0)[l]{$\sq_j^\b$}}
\put(101.5,163){\makebox(0,0)[r]{$g$}}
\put(99,140){\makebox(0,0)[t]{\bf{(d)}}}


\put(11,111){\makebox(0,0)[r]{$\sq_i^\a$}}
\put(48,125){\makebox(0,0)[l]{$\sq_j^\b$}}
\put(48,96){\makebox(0,0)[l]{$V$}}
\put(35,117){\makebox(0,0)[r]{$\sg$}}
\put(35,105){\makebox(0,0)[r]{$q^\a$}}
\put(41,111.5){\makebox(0,0)[l]{$q^\b$}}
\put(28,96){\makebox(0,0)[t]{\bf{(e)}}}

\put(68,111){\makebox(0,0)[r]{$\sq_i^\a$}}
\put(105,125){\makebox(0,0)[l]{$\sq_j^\b$}}
\put(105,96){\makebox(0,0)[l]{$V$}}
\put(87,104){\makebox(0,0)[r]{$\sq_n^\a$}}
\put(98.5,111.5){\makebox(0,0)[l]{$\sq_m^\b$}}
\put(87,96){\makebox(0,0)[t]{\bf{(f)}}}


\put(3.5,68){\makebox(0,0)[r]{$\sq_i$}}
\put(16.5,66.5){\makebox(0,0)[t]{$\sq_i$}}
\put(17,75){\makebox(0,0)[b]{$g$}}
\put(29,68){\makebox(0,0)[l]{$\sq_i$}}
\put(16,60){\makebox(0,0)[t]{\bf{(g)}}}
\put(45,68){\makebox(0,0)[r]{$\sq_i$}}
\put(58,66){\makebox(0,0)[t]{$q$}}
\put(58,74.5){\makebox(0,0)[b]{$\sg$}}
\put(71,68){\makebox(0,0)[l]{$\sq_n^{}$}}
\put(58,60){\makebox(0,0)[t]{\bf{(h)}}}
\put(86,68){\makebox(0,0)[r]{$\sq_i$}}
\put(99,79){\makebox(0,0)[b]{$\sq_{1,2}$}}
\put(112,68){\makebox(0,0)[l]{$\sq_k^{}$}}
\put(99,60){\makebox(0,0)[t]{\bf{(i)}}}


\put(-0.5,30.5){\makebox(0,0)[r]{$\sq_i^\a$}}
\put(35,45){\makebox(0,0)[l]{$\sq_j^\b$}}
\put(35,16){\makebox(0,0)[l]{$V$}}
\put(18.5,38){\makebox(0,0)[b]{$g$}}
\put(16,15){\makebox(0,0)[t]{\bf{(j)}}}
\put(40,30.5){\makebox(0,0)[r]{$\sq_i^\a$}}
\put(76,45){\makebox(0,0)[l]{$\sq_j^\b$}}
\put(76,16){\makebox(0,0)[l]{$V$}}
\put(76,37){\makebox(0,0)[l]{$g$}}
\put(57,15){\makebox(0,0)[t]{\bf{(k)}}}
\put(81.5,30.5){\makebox(0,0)[r]{$\sq_i^\a$}}
\put(117.5,45){\makebox(0,0)[l]{$\sq_j^\b$}}
\put(117.5,16){\makebox(0,0)[l]{$V$}}
\put(117.5,30){\makebox(0,0)[l]{$g$}}
\put(99,15){\makebox(0,0)[t]{\bf{(\boldmath{$\ell$})}}}
\end{picture}
\begin{minipage}[b]{13cm}
  {\bf Fig.~1:} 
  Feynman diagrams relevant for the ${\cal O}(\a_{s})$ SUSY-QCD 
  corrections to squark decays into vector bosons. 
\end{minipage}


\noi
\vspace*{5mm}
\begin{minipage}[b]{130mm} 
\begin{picture}(130,80)
\put(4,4){\mbox{\epsfig{figure=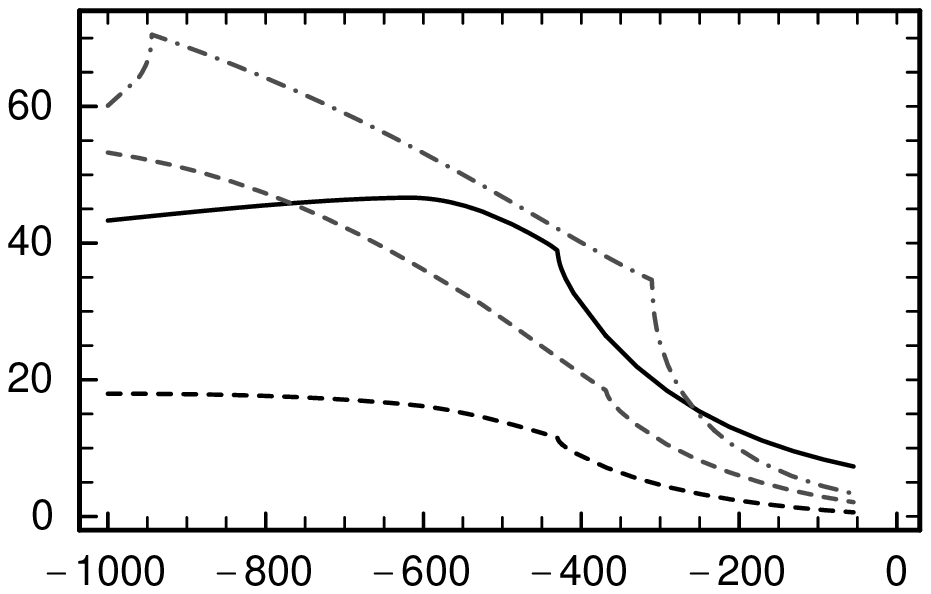,height=6.6cm}}}
\put(5,33){\makebox(0,0)[br]{{\rotatebox{90}{\large BR~[\%]}}}}
\put(63,2){\makebox(0,0)[tc]{{\large $\mu$~[GeV]}}}
\put(105,60){\makebox(0,0)[br]{{\large \bf{(a)}}}}
\put(105,52){\makebox(0,0)[br]{{\boldmath{$\tan\b=2$}}}}
\put(49,59){\makebox(0,0)[bl]{{$\sb_1\to\st_1 W^-$}}}
\put(24,54){\makebox(0,0)[bl]{{$\sb_2\to\st_1 W^-$}}}
\put(22,45){\makebox(0,0)[tl]{{$\st_2\to\st_1 Z^0$}}}
\put(30,29){\makebox(0,0)[bl]{{$\st_2\to\sb_1 W^+$}}}
\end{picture} 
\vspace{3mm}\\
\begin{picture}(130,80)
\put(6,4){\mbox{\epsfig{figure=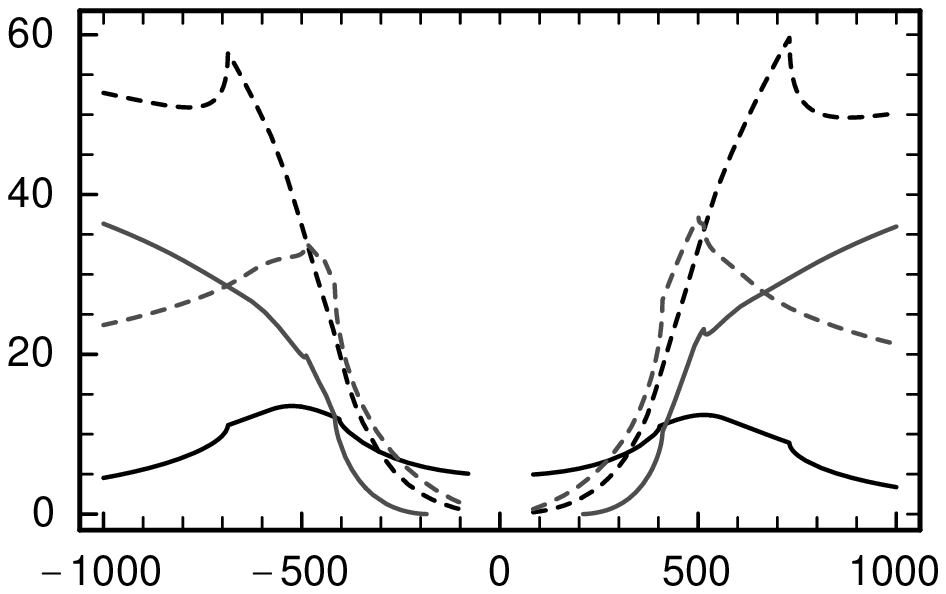,height=6.6cm}}}
\put(5,33){\makebox(0,0)[br]{{\rotatebox{90}{\large BR~[\%]}}}}
\put(63.5,2){\makebox(0,0)[tc]{{\large $\mu$~[GeV]}}}
\put(106,62){\makebox(0,0)[br]{{\large \bf{(b)}}}}
\put(63,63){\makebox(0,0)[bc]{{\boldmath{$\tan\b=40$}}}}
\put(53.5,55){\makebox(0,0)[bl]{{$\st_2\to\sb_1 W^+$}}}
\put(51,56){\vector(-3,-1){11}}
\put(75,56){\vector(3,-1){13}}
\put(53.5,46){\makebox(0,0)[bl]{{$\sb_2\to\st_1 W^-$}}}
\put(51,47){\vector(-3,-2){6}}
\put(75,47){\vector(3,-1){8}}
\put(53.5,37){\makebox(0,0)[bl]{{$\sb_2\to\sb_1 Z^0$}}}
\put(51,38){\vector(-3,-1){11}}
\put(73,38){\vector(3,-1){13}}
\put(53.5,28){\makebox(0,0)[bl]{{$\st_2\to\st_1 Z^0$}}}
\put(51,29){\vector(-3,-1){9}}
\put(73,29){\vector(3,-1){12}}
\end{picture}
\vspace{2mm}\\
{\bf Fig.~2:} Tree--level branching ratios for stop and sbottom decays 
into vector bosons as a function of $\mu$, for 
$M_{\ti Q}=500$ GeV, $M_{\ti U}=444$ GeV, $M_{\ti D}=556$ GeV, 
$A_t=A_b=500$ GeV, $M=200$ GeV, and $m_A=200$ GeV. 
{\bf (a)} $\tan\b = 2$, and {\bf (b)} $\tan\b = 40$.
\end{minipage}

\clearpage
\noi
\begin{minipage}[b]{130mm} 
\begin{picture}(130,85)
\put(6,4){\mbox{\epsfig{figure=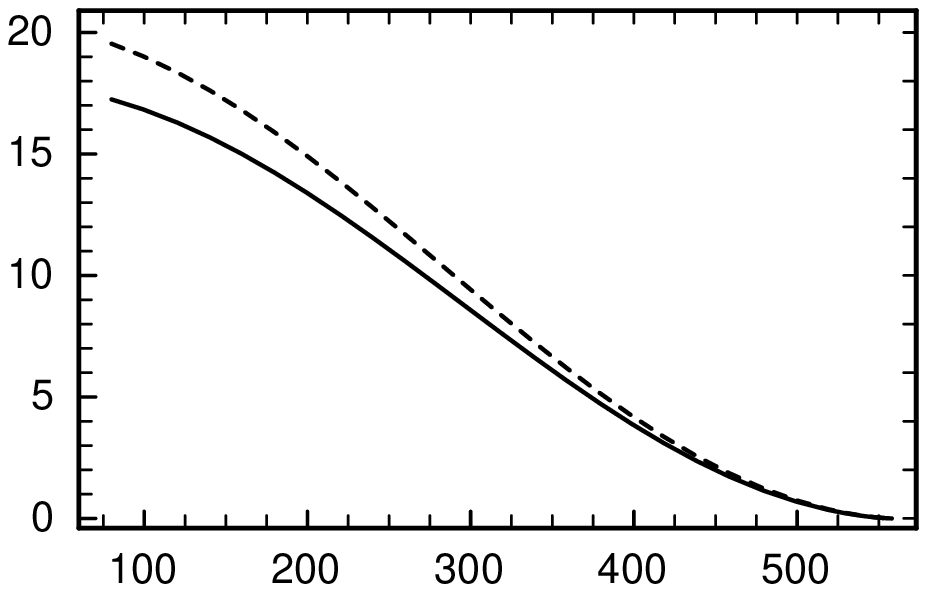,height=6.6cm}}}
\put(7,32.5){\makebox(0,0)[br]{{\rotatebox{90}{\large $\Gamma$~[GeV]}}}}
\put(65,2){\makebox(0,0)[tc]{{\large $\mst{1}$~[GeV]}}}
\end{picture}
\vspace{1mm}\\
{\bf Fig.~3:}
  Tree--level (dashed line) and SUSY--QCD corrected (solid line) 
  widths of the decay $\st_2\to \st_1 Z^0$ as a function 
  of $\mst{1}$ for $\mst{2} = 650\gev$, $\cst = -0.6$, and 
  $\msg=500\gev$.
\end{minipage}


\noi
\begin{minipage}[b]{130mm} 
\vspace*{-3mm}
\begin{picture}(130,80)
\put(6,2){\mbox{\epsfig{figure=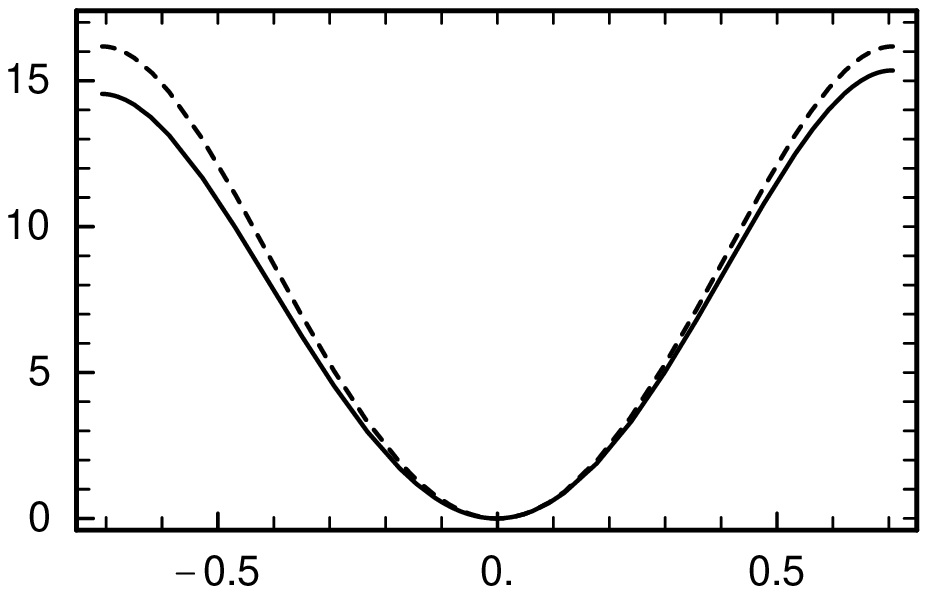,height=6.6cm}}}
\put(7,30){\makebox(0,0)[br]{{\rotatebox{90}{\large $\Gamma$~[GeV]}}}}
\put(66,1){\makebox(0,0)[tc]{{\large $\cst$}}}
\end{picture}
\vspace{1mm}\\
{\bf Fig.~4a:}
  Tree--level (dashed line) and SUSY--QCD corrected (solid line) 
  widths of the decay $\st_2\to \st_1 Z^0$ as a function 
  of $\cst$ for $\mst{1}=200\gev$, $\mst{2} = 650\gev$,  
  and $\msg=500\gev$. \\
\vspace{5mm}
\begin{picture}(130,80)
\put(6,2){\mbox{\epsfig{figure=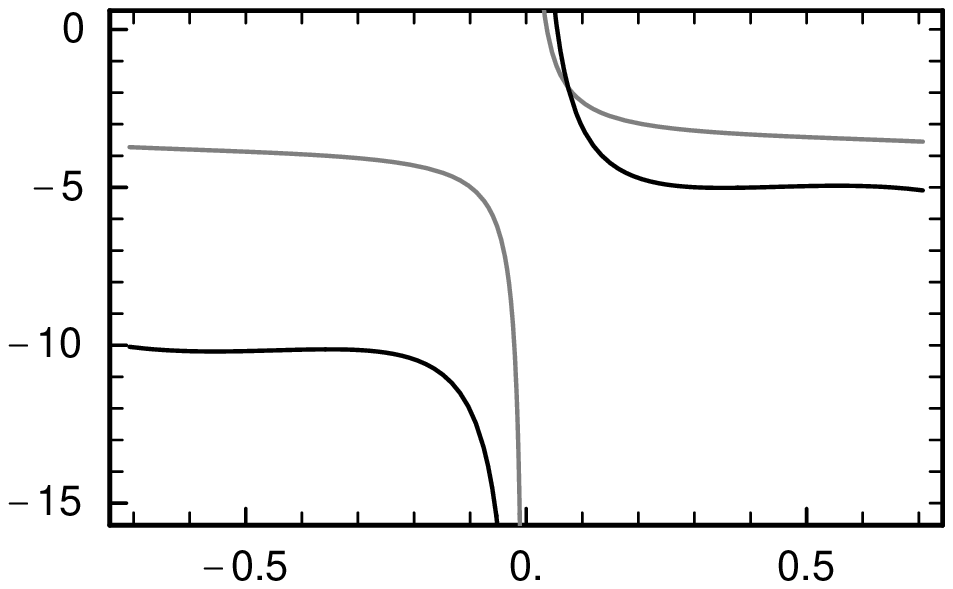,height=6.6cm}}}
\put(7,28){\makebox(0,0)[br]{{\rotatebox{90}{\large $\d\G/\G^0$~[\%]}}}}
\put(67,1){\makebox(0,0)[tc]{{\large $\cst$}}}
\end{picture}
\vspace{1mm}\\
{\bf Fig.~4b:}
  SUSY--QCD correction relative to the tree--level  
  width of the decay $\st_2\to \st_1 Z^0$ as a function 
  of $\cst$ for $\mst{1}=200$ GeV and $\mst{2} = 650$ GeV.  
  The black (gray) lines are for $\msg=500$ (1000) GeV.
\end{minipage}


\noi
\begin{minipage}[b]{130mm} 
\vspace*{-5mm}
\begin{picture}(130,80)
\put(6,4){\mbox{\epsfig{figure=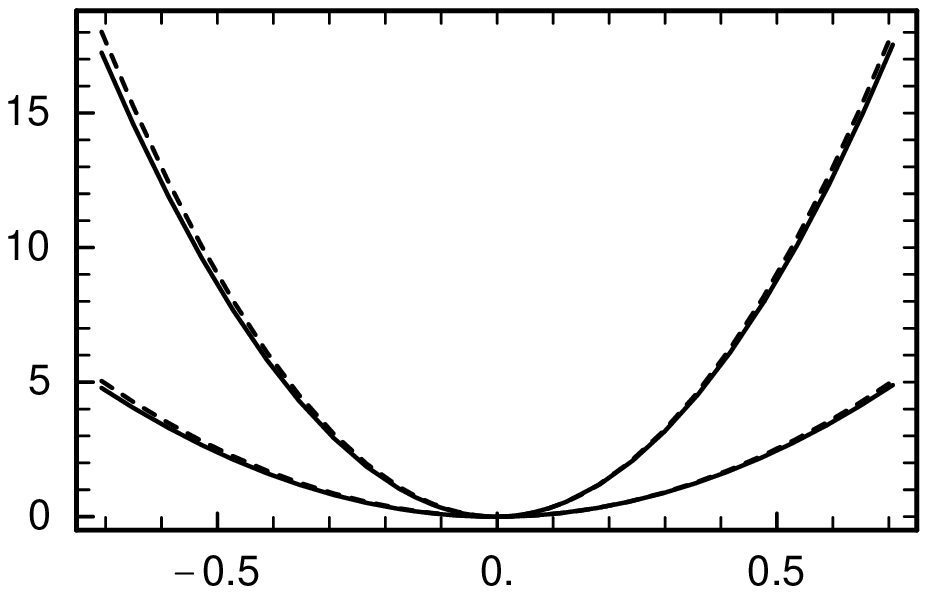,height=6.6cm}}}
\put(7,32){\makebox(0,0)[br]{{\rotatebox{90}{\large $\Gamma$~[GeV]}}}}
\put(66,2){\makebox(0,0)[tc]{{\large $\cst$}}}
\put(32,47){\makebox(0,0)[bl]{{\large \boldmath{$\sb_1$}}}}
\put(27,26){\makebox(0,0)[bl]{{\large \boldmath{$\sb_2$}}}}
\end{picture}
\vspace{0.1mm} \\
{\bf Fig.~5a:}
  Tree--level (dashed lines) and SUSY--QCD corrected (solid lines) 
  widths of the decays $\sb_{1,2}\to \st_1 W^-$ as a function 
  of $\cst$ for $\msb{1}=500\gev$, $\msb{2} = 520\gev$, $\csb = -0.9$, 
  $\mst{1}=200\gev$, $\msg=520\gev$, and $\tan\b=2$. $\mst{2}$ is 
  a function of the other parameters and varies with $\cst$.\\
\vspace{1mm}
\begin{picture}(130,80)
\put(6,4){\mbox{\epsfig{figure=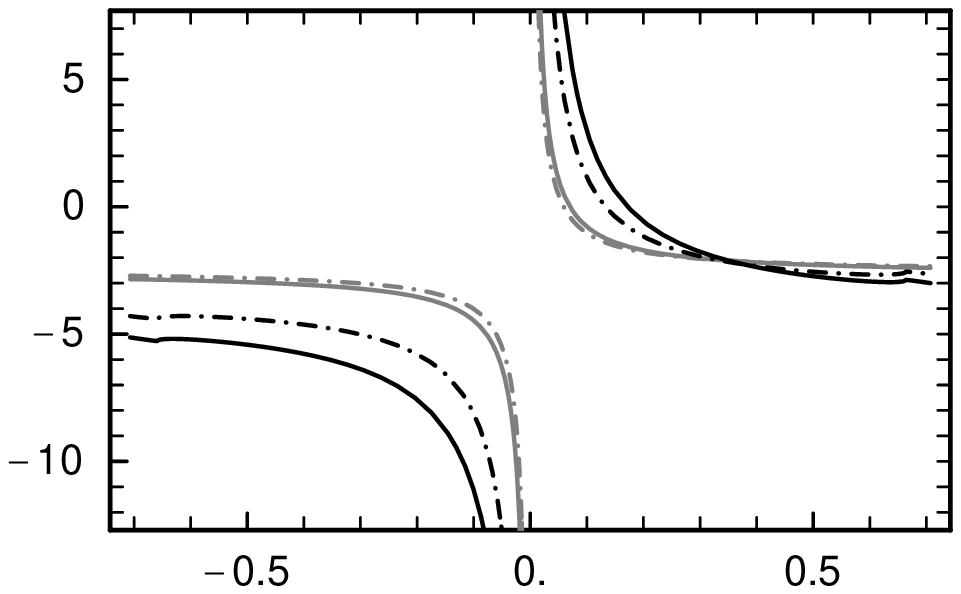,height=6.6cm}}}
\put(7,31){\makebox(0,0)[br]{{\rotatebox{90}{\large $\d\G/\G^0$~[\%]}}}}
\put(67,2){\makebox(0,0)[tc]{{\large $\cst$}}}
\end{picture}
\vspace{0.1mm} \\
{\bf Fig.~5b:}
  SUSY--QCD corrections relative to the tree--level widths of the decays 
  $\sb_{1}\to \st_1 W^-$ (dashdotted lines) and 
  $\sb_{2}\to \st_1 W^-$ (solid lines) as a function of $\cst$ 
  for $\msb{1}=500\gev$, $\msb{2} = 520\gev$, $\csb = -0.9$, 
  $\mst{1}=200\gev$, and $\tan\b=2$. The black (gray) lines are for 
  $\msg=520$ (1000) GeV.
  $\mst{2}$ is a function of the other parameters, as in Fig.~5a.\\
\end{minipage}


\noi
\begin{minipage}[b]{130mm} 
\vspace*{-5mm}
\begin{picture}(130,80)
\put(6,4){\mbox{\epsfig{figure=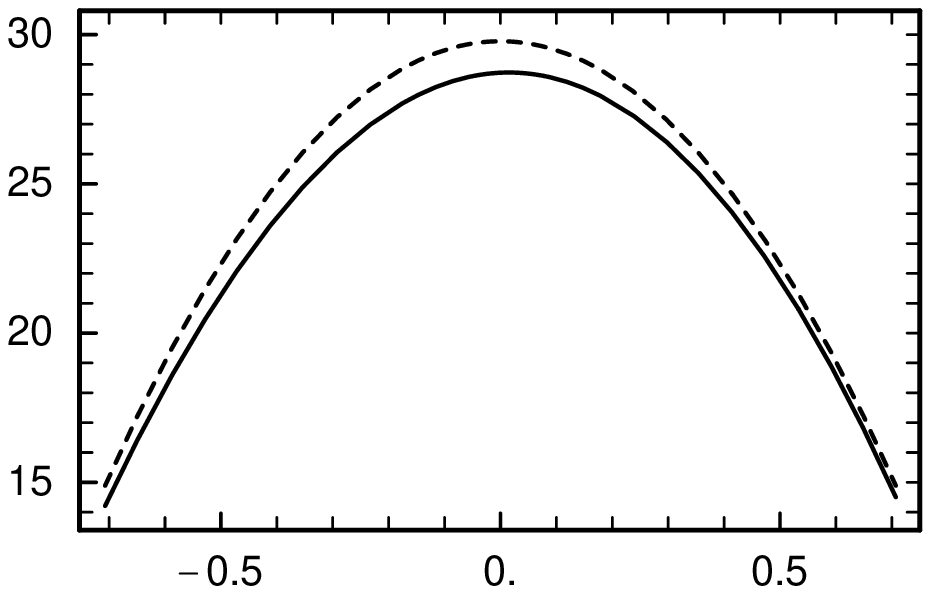,height=6.6cm}}}
\put(7,32){\makebox(0,0)[br]{{\rotatebox{90}{\large $\Gamma$~[GeV]}}}}
\put(66,2){\makebox(0,0)[tc]{{\large $\cst$}}}
\end{picture}  
\vspace{0.1mm} \\
{\bf Fig.~6a:}
  Tree--level (dashed line) and SUSY-QCD corrected (solid line) 
  widths of the decay $\st_2\to \sb_1 W^+$ as a function 
  of $\cst$ for $\mst{1}=300\gev$, $\mst{2} = 650\gev$, 
  $\msb{1}=380\gev$, $\csb = -0.8$, $\msg=500\gev$, and $\tan\b=40$. 
  $\msb{2}$ is a function of the other parameters and varies 
  with $\cst$.\\
\vspace{1mm}
\begin{picture}(130,80)
\put(6,4){\mbox{\epsfig{figure=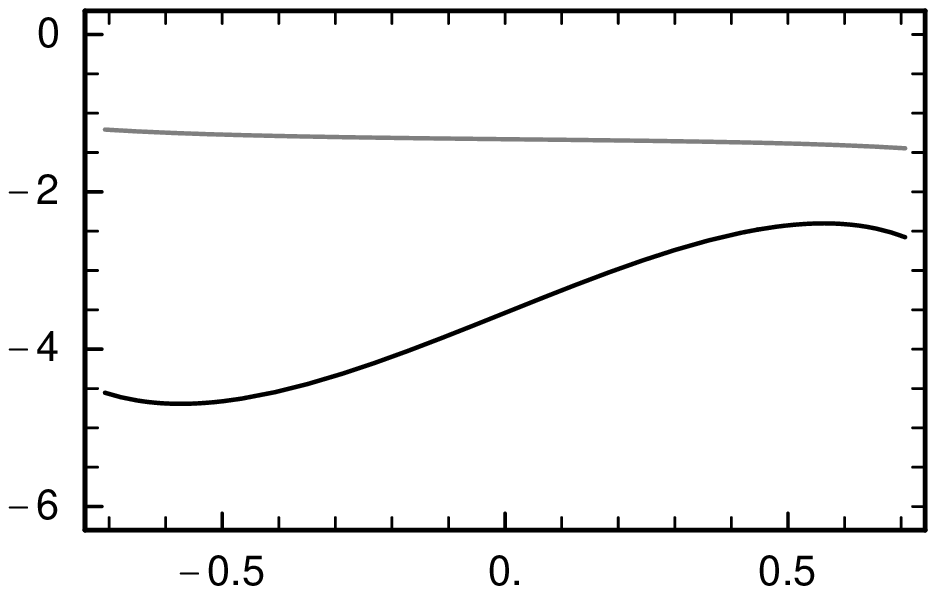,height=6.6cm}}}
\put(7,30){\makebox(0,0)[br]{{\rotatebox{90}{\large $\d\G/\G^0$~[\%]}}}}
\put(66,2){\makebox(0,0)[tc]{{\large $\cst$}}}
\end{picture} 
\vspace{0.1mm} \\
{\bf Fig.~6b:}
  SUSY--QCD corrections relative to the tree--level width of the 
  decay $\st_{2}\to \sb_1 W^+$ as a function 
  of $\cst$ for $\mst{1}=300\gev$, $\mst{2}=650\gev$, 
  $\msb{1}=380\gev$, $\csb = -0.8$, and $\tan\b=40$. 
  The black (gray) line is for $\msg=500$ (1000) GeV. 
  $\msb{2}$ is a function of the other parameters, as in Fig.~6a.\\
\end{minipage}

\end{center}

\baselineskip=14pt   

\begin{thebibliography}{99}

\bibitem{haber}  
For a review, see: \\ 
H. P. Nilles, Phys.~Rep. 110 (1984) 1;\\
H.E. Haber, G.L. Kane, Phys. Rep. 117 (1985) 75;\\
R. Barbieri, Riv. Nuov. Cim. 11 (1988) 1.

\bibitem{ellis} 
J. Ellis, S. Rudaz, Phys. Lett. B128 (1983) 248; \\
J.F. Gunion, H.E. Haber, Nucl. Phys. B272 (1986) 1; B402 (1993) 567 (E). 

\bibitem{hikasa}
K. Hikasa, M. Kobayashi, Phys. Rev. D36 (1987) 724.

\bibitem{drees} 
M. Drees, K. Hikasa, Phys. Lett. B252 (1990) 127. 

\bibitem{beenakker} 
W. Beenakker, R. H\"opker, P. M. Zerwas, 
Phys. Lett. B349 (1995) 463.

\bibitem{djouadi}
A. Arhrib, M. Capdequi--Peyranere, A. Djouadi, Phys. Rev. D52 (1995) 1404.

\bibitem{helmut}
H. Eberl, A. Bartl, W. Majerotto, 
Nucl. Phys. B472 (1996) 481.

\bibitem{hadro} 
W. Beenakker, R. H\"opker, M. Spira, P. M. Zerwas, 
Nucl. Phys. B492 (1997) 51.

\bibitem{porod} 
A. Bartl, W. Majerotto, W. Porod, Z. Phys. C 64 (1994) 499; 
C 68 (1995) 518 (E).

\bibitem{nlc}
A. Bartl, H. Eberl, S. Kraml, W. Majerotto, W. Porod, A. Sopczak, 
  hep-ph/9701336, to appear in Z. Phys. C.

\bibitem{ecfa}
ECFA/DESY LC Physics Working Group: E. Accomando, et al., 
  DESY 97--100, hep-ph/9705442, to be published in Phys. Rep.; \\
A. Bartl, H. Eberl, T. Gajdosik, S. Kraml, W. Majerotto, 
W. Porod, A. Sopczak, hep-ph/9709252, contribution to the proceedings   
``ECFA/DESY Study on Physics and Detectors for the Linear Collider'', 
DESY 97--123E, ed. R. Settles. 

\bibitem{chnt}
S. Kraml, H.Eberl, A. Bartl, W. Majerotto, W. Porod, 
Phys. Lett. B386 (1996) 175; \\
A. Djouadi, W. Hollik, C. J\"unger, 
Phys. Rev. D55 (1997) 6975.

\bibitem{higgsdec}
A. Arhrib, A. Djouadi, W. Hollik, C. J\"unger, hep-ph/9702426.

\bibitem{gluino}
W. Beenakker, R. H\"opker, P. M. Zerwas, 
Phys. Lett. B378 (1996) 159; \\
W. Beenakker, R. H\"opker, T. Plehn, P. M. Zerwas, 
Z. Phys. C 75 (1997) 349.

\bibitem{siegel}
W. Siegel, Phys. Lett. B84 (1979) 193; \\ 
D. M. Capper, D. R. T. Jones, P. van Nieuwenhuizen, 
Nucl. Phys. B167 (1980) 479; \\
I. Jack, D. R. T. Jones, hep-ph/9707278.

\bibitem{pave}
G. Passarino, M. Veltman, Nucl. Phys. B160 (1979) 151.

\bibitem{denner}
A. Denner, Fortschr. Phys. 41 (1993) 307. 

\bibitem{mQproblem}
A. Bartl, H. Eberl, K. Hidaka, T. Kon, W. Majerotto, Y. Yamada, 
Phys. Lett. B402 (1997) 303. 

\end{thebibliography}
\end{document}